\documentclass[aip,jcp,numerical,reprint,longbibliography,groupedaddress]{revtex4-1}
\usepackage[utf8]{inputenc}
\usepackage{graphicx}
\usepackage{amssymb, amsmath}
\usepackage{color,soul}

\begin{document}
\title{Coarsening and Aging of Lattice Polymers: Influence of Bond Fluctuations}
 
\author{Henrik Christiansen}
\email{henrik.christiansen@itp.uni-leipzig.de}
\author{Suman Majumder}
\email{suman.majumder@itp.uni-leipzig.de}
\author{Wolfhard Janke}
\email{wolfhard.janke@itp.uni-leipzig.de}
\affiliation{Institut für Theoretische Physik, Universität Leipzig, Postfach 100 920, 04009 Leipzig, Germany}
\date{\today}

 \begin{abstract}
We present results for the nonequilibrium dynamics of collapse for a model flexible homopolymer on simple cubic lattices with fixed and fluctuating bonds between the monomers. 
Results from our Monte Carlo simulations show that, phenomenologically, the sequence of events observed during the collapse are independent of the bond criterion. While 
the growth of the clusters (of monomers) at different temperatures exhibits a nonuniversal power-law behavior when the bonds are fixed, 
the introduction of fluctuations in the bonds by considering the existence of diagonal bonds produces a temperature independent 
growth, which can be described by a universal nonequilibrium finite-size scaling function with a non-universal metric 
factor. We also examine the related aging phenomenon, probed by a suitable two-time density-density autocorrelation function showing
a simple power-law scaling with respect to the growing cluster size. 
Unlike the cluster-growth exponent $\alpha_c $, the nonequilibrium autocorrelation exponent $\lambda_C$ governing the aging during the collapse, 
however, is independent of the bond type and strictly follows the bounds proposed by two of us in \textit{Phys. Rev. E} \textbf{93}, 032506 (2016) at all temperatures.
\end{abstract}

\maketitle
\section{Introduction}
Despite their apparent extreme simplification, lattice models have been proved to be very handy in Monte Carlo (MC) simulations \cite{landau2014guide} 
providing useful insights in problems related to various phase transitions, e.g., gas-liquid transition,\cite{Lattice_gas} ferromagnetic transition,\cite{baxter1982exactly} 
collapse transition of a polymer,\cite{grassberger1995simulations,grassberger_perm,vogel2007freezing} etc. 
In particular, the behavior of the static properties obtained from such simulations of lattice models have been found to be in fairly good agreement with 
the corresponding theories and often with experiments. However, the dynamic properties seem to be largely dependent on the choice of the 
model as well the implemented ``local moves''. In this paper, we aim to understand similar effects in the nonequilibrium dynamics of the collapse 
transition of lattice polymers. 
\par
Collapse transition refers to the change in conformation a polymer chain, initially in an expanded coil state under good solvent conditions (or high temperature), 
experiences when transferred to a bad solvent (low temperature), where the equilibrium conformation is globular in nature.\cite{stockmayer1960problems,nishio1979first}
The significance of such collapse transition lies in its close association with the folding process of certain macromolecules like 
proteins.\cite{camacho1993kinetics,haran2012} Ever since its introduction, self-avoiding walks (SAW) on lattices 
\cite{domb2009self} have been successfully used to obtain the critical exponent $\nu$ related to the size of the polymer,\cite{flory1953principles} i.e., 
the radius of gyration $R_g \sim N^\nu$, where $N$ is the number of monomers in the 
polymer chain. Furthermore, by introducing an attractive interaction for the nearest-neighbor non-bonded contacts, and conducting MC simulations of such interactive self-avoiding walk (ISAW) on lattices,
one can capture the collapse transition as well as the freezing or crystallization.\cite{rampf2005first,vogel2007freezing} Thus, on the one hand, 
the static properties of a polymer in equilibrium have been quite well studied using lattice as well as off-lattice models. 
On the other hand, the dynamic properties have received less attention. For dynamic studies of polymers, there is a quest \cite{verdier1962monte} for a suitable model and appropriate set of moves that reproduces the well-known 
Rouse dynamics in equilibrium,\cite{rouse1953theory} valid in absence of hydrodynamics. These led to the introduction of bond-fluctuation models \cite{carmesin1988bond} 
and diagonal bond models \cite{shaffer1994effects,dotera1996diagonal} which not only reproduce the static properties correctly but also provide reliable dynamics. 
However, the application of such lattice models to understand the nonequilibrium kinetics of the collapse transition has rarely been attempted.
\par
Current developments in experimental techniques have made it a lot easier to monitor a single polymer,\cite{tang2011compression,tress2013glassy} in turn urging more interests in 
the dynamics of a single polymer via computer simulations. In this regard, one can 
understand the collapse dynamics of a homopolymer \cite{majumder2015cluster,majumder2016evidence,majumder2017kinetics} by drawing analogies with 
usual nonequilibrium coarsening phenomena of particle or spin systems.\cite{Bray2002,puri2004kinetics}
Especially, the scaling of the growth of monomer clusters formed during the collapse and the scaling of 
the two-time density-density autocorrelation functions (showing aging) are worth mentioning. Phenomenologically, the events that occur during the collapse 
can be well described by the ``pear-necklace'' picture of Halperin and Goldbart (HG).\cite{HG_pearlnecklace} In accordance with HG, confirmed both in lattice \cite{kuznetsov1995kinetics} 
and off-lattice simulations (both without hydrodynamics \cite{majumder2015cluster,majumder2017kinetics} and with hydrodynamics \cite{Abrams2002,Yeomans2005}),
the polymer collapse starts with the formation of small clusters along the chain at locally higher densities. 
Those clusters subsequently become stable and start to coarsen by accumulating monomers from the connecting bridges. Once those bridges stiffen, 
clusters start to coalesce with each other until only a single globular cluster is left. Finally, the monomers within the globule rearrange to form an even more compact configuration, 
minimizing the surface energy. The growth of the clusters during the coarsening or coalescence phase of the collapse can be viewed under the light of 
well known ordering or coarsening kinetics. In earlier studies the cluster growth was shown to follow a simple power law:
$C_s(t) \sim t^{\alpha_c}$ (where $C_s(t)$ is the cluster size at time $t$ ) and the cluster-growth exponent $\alpha_c$ was found to be $1/2$ for lattice polymers \cite{kuznetsov1995kinetics}
and $\alpha_c=2/3$ for an off-lattice model,\cite{byrne1995kinetics} consistent with a Gaussian self-consistent theory.\cite{kuznetsov1996kinetic}
However, recently, in an off-lattice model \cite{majumder2015cluster,majumder2017kinetics} with diffusive dynamics 
(in absence of hydrodynamics) it has been shown that the average cluster size, $C_s(t)$, obeys a scaling of the form 
\begin{equation}\label{cluster_growth}
 C_s(t)=C_0+At^{\alpha_c},
\end{equation}
where $C_0$ is the crossover (from the initial cluster-formation stage to the coarsening stage) cluster size. 
The corresponding growth exponent for this model is $\alpha_c \approx 1$, as observed for Ostwald ripening.\footnote{This corresponds to the familiar value of the Ostwald ripening exponent $1/3$ obtained when considering the linear size of ordered structures.}
Moreover, using the scaling form \eqref{cluster_growth} of the cluster growth it has been shown that $\alpha_c$ is independent of the quench temperatures and the growth 
can be described by a universal nonequilibrium finite-size scaling function with a non-universal metric factor.\cite{majumder2017kinetics}
\par
In analogy with the nonequilibrium ordering or coarsening processes of particle or spin systems, 
another intriguing feature observed during the collapse is the presence of aging,\cite{zannetti_book,henkel2010non} 
generally probed by a two-time density-density autocorrelation function $C(t,t_w)$ (where $t$ is the observation time 
and $t_w$ is the waiting time). In this context one is particularly interested in 
the related dynamic scaling given as 
\begin{equation}\label{aging_scaling}
C(t,t_w) =A_Cx_{\!_C}^{-\lambda_C};~x_{\!_C}= \frac{C_s(t)}{C_s(t_w)}.
\end{equation}
Such power-law scaling is reminiscent of the scaling observed in the Ising model with both nonconserved\cite{fisher1988nonequilibrium,humayun1991non,liu1991nonequilibrium,henkel2001aging,henkel2004two,lorenz2007numerical,midya2014aging} and conserved \cite{midya2015dimensionality} order parameter,
where the two-time order-parameter autocorrelation function scales with the ratio of the length scales at the concerned times.
In Refs.\ \onlinecite{majumder2016evidence,majumder2017kinetics}, 
for collapse in an off-lattice model, it has been shown that the value of the nonequilibrium autocorrelation exponent $\lambda_C \approx 1.25$ in \eqref{aging_scaling} 
is independent of the quench temperature and obeys a theoretically predicted bound \cite{majumder2016evidence} 
\begin{eqnarray}\label{poly-bound}
 (\nu d-1) \le \lambda_C \le 2(\nu d-1), 
\end{eqnarray}
where $d$ is the dimension and $\nu$ is the previously discussed static critical exponent related to the size of the polymer. 
Such a dimension-dependent bound was first proposed for aging in ordering spin glasses \cite{fisher1988nonequilibrium} and later verified for spin systems having nonconserved order-parameter dynamics. \cite{liu1991nonequilibrium}
A more general and in fact sharpened bound that also includes the conserved order-parameter dynamics case was later proposed in Ref.\ \onlinecite{yeung1996bounds}.
The bound \eqref{poly-bound}, too, is general for the collapse of a polymer and, although not yet verified, expected to be valid in presence of hydrodynamics as well.
\par
In spite of the simplicity of implementation, the investigation of the above mentioned scaling laws related to the nonequilibrium dynamics of collapse transition 
using a lattice model has been ignored so far. Motivated by that, we present comparative results from MC simulations of 
two different lattice models with fixed and fluctuating bonds. With the primary focus on the various scaling laws related to the collapse, 
we show that while the model with fixed bonds does not provide a universal picture of the cluster growth, the model with fluctuating bonds 
yields a scaling independent of the quench temperature. However, the scaling \eqref{aging_scaling} related to aging is independent of the models considered, indicating a dynamic universal behavior of aging.
\par
The rest of the paper is organized in the following manner. Next, in Section \ref{Model}, we describe the models and the method 
of simulation used. Then, in Section \ref{results}, we present our main results concerning the cluster growth and aging
followed by a discussion and conclusion in Section \ref{conclusion}.

\section{Models and Method}\label{Model}
For our polymer model we consider an ISAW on a simple cubic lattice with unit lattice constant, fixing the unit of length. In this model each lattice site can be occupied by a single monomer. 
The energies leading to the collapse transition are governed by the Hamiltonian
\begin{equation}\label{hamiltonian}
H=-\frac{1}{2} \sum_{i \ne j,  j \pm 1} w(r_{ij}),
\end{equation}
where $i$ and $j$ correspond to a non-bonded pair of monomers, $r_{ij}$ is the Euclidean distance between them, and $w(r_{ij})$ is the distance-dependent interaction parameter. 
We use the simplest case of nearest-neighbor interaction as
\begin{equation}\label{interaction}
  w(r_{ij})=\begin{cases} \epsilon & r_{ij} = 1 \\ 0 & \text{else}\end{cases}.
\end{equation}
For computational convenience and comparability to previous works \cite{vogel2007freezing,bachmann2004thermodynamics} we choose $\epsilon=1$, which sets the energy respectively the temperature scales 
(the unit of temperature is $\epsilon/k_B$, where the Boltzmann factor $k_B$ is set to unity).
The Hamiltonian given by \eqref{hamiltonian} and \eqref{interaction}
favors at low temperature more and more nearest-neighbor non-bonded contacts, thus facilitating a coil-globule transition 
in the model. For our studies, we use two different criteria for the bonds connecting the adjacent monomers. In one case, we fix the bond distances to $1$ which from now on we refer to as Model\ I. 
In the other case we allow a fluctuation in the bond length by additionally considering diagonal bonds, i.e., there the possible bond lengths are $1$, $\sqrt{2}$ and $\sqrt{3}$. This we refer to as Model\ II.
Note that in both cases the attractive interaction \eqref{interaction} only acts between monomers located at nearest-neighbor sites.
The thermodynamic properties of Model\ I are well studied for both the freezing and collapse transition.\cite{vogel2007freezing} Model\ II has its origin from the bond-fluctuation model 
of Carmesin and Kremer \cite{carmesin1988bond} and has been independently studied both for the static and dynamic properties in equilibrium.\cite{shaffer1994effects,dotera1996diagonal,subramanian2008}
\par
We introduce the dynamics in the models via Markov chain Monte Carlo simulation. 
For Model\ I a trial move of a randomly picked monomer in the chain could be a end-move, corner-move, or the crankshaft-move, depending on the position of the monomer. Care has 
to be taken to preserve the excluded volume condition (no two monomers can occupy the same lattice site) and to maintain the bond connectivity between adjacent monomers.
Since in Model\ II we allow a fluctuation in the bond length, a trial move only consists of local displacement of a randomly picked monomer with the 
constraint of preserving the excluded volume condition and the bond connectivity. For details on the allowed moves in Model\ II we refer to Ref.\ \onlinecite{dotera1996diagonal}. 
For both the models a trial move is accepted or rejected following the Metropolis algorithm with Boltzmann criterion.
A single Monte Carlo step (MCS) consists of $N$ (where $N$ is the number of monomers in the chain) attempted moves on randomly picked monomers, 
effectively setting the time scale.  
\par
The thermodynamic collapse transition temperature is $T_{\theta}(N \rightarrow \infty) \approx 3.7$ in Model\ I.\cite{vogel2007freezing} 
For Model\ II there is no study available that quantifies $T_{\theta}$. We therefore first performed a set of equilibrium simulations and using multiple-histogram reweighting \cite{ferrenberg1989optimized,janke2003histograms}
obtained an estimate of $T_{\theta}(4096)\approx 4.0$ for Model\ II,
which is comparatively crude but serves the purpose of indicating the relevant temperature range.
For both the models, we prepare well equilibrated initial configurations at high temperatures $T_h=6\approx 1.5T_{\theta}$ that mimics an extended coil polymer and then quench it 
to the globular phase at different temperatures $T_q < T_{\theta}$. We use chains of length $N$ within a wide range ($512 \le N \le 8192$).
By using lattice polymers we are thereby able to simulate polymers one order of magnitude longer
than in the off-lattice simulations performed in Refs.\ \onlinecite{majumder2015cluster,majumder2016evidence,majumder2017kinetics}.
\par
The time evolution of a single simulation run does depend of course on the randomly chosen initial polymer configuration in the high-temperature extended coil phase. 
To arrive at meaningful results, the data presented are hence averaged over $300$ different initial realizations. 

\section{Nonequilibrium Dynamics of the Collapse Transition}
\label{results}
\begin{figure}[!ht]
\includegraphics{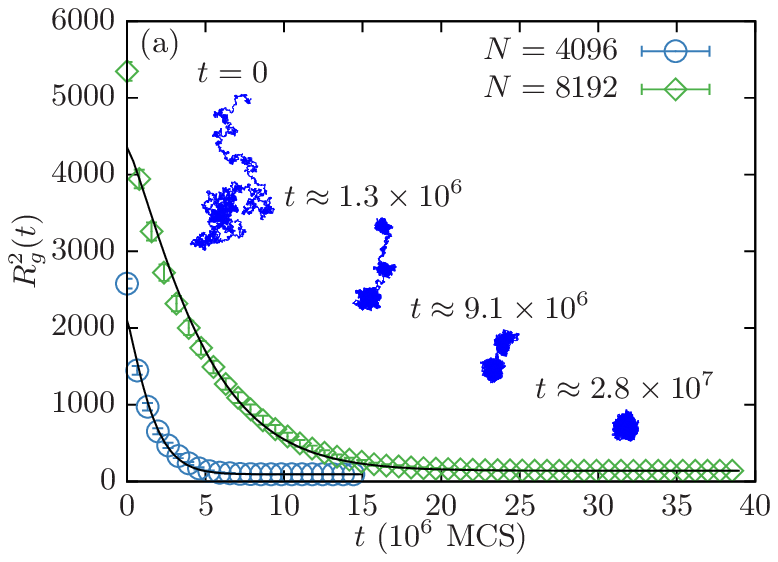}
\includegraphics{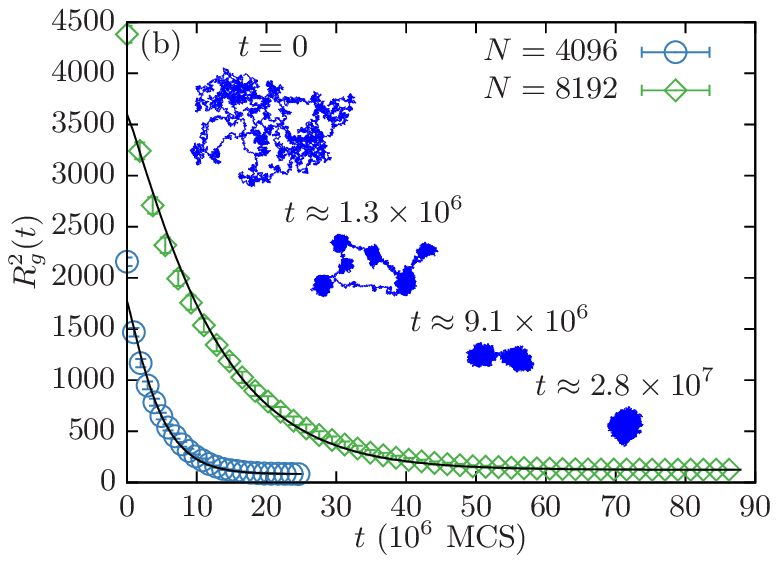}
\caption{The decay of the squared radius of gyration for (a) Model\ I with fixed bonds and (b) Model\ II with fluctuating bonds at $T_q=2.5$ for polymers with $N=4096$ and $8192$. 
The solid black lines show the results of the fit to Eq.\ \eqref{EqFit}.
Additionally we have included for both models exemplary evolution snapshots of polymer conformations with $N=8192$.}
\label{R_g}
\end{figure}
We now continue with the results and analyses of the nonequilibrium dynamics of the collapse.
In Figs.\ \ref{R_g}(a) and (b), we show the time evolution snapshots of a polymer quenched to $T_q=2.5$, for both the models with $N=8192$.
Chronologically, the observed events in both the models are in good accordance with the phenomenological picture of HG.\cite{HG_pearlnecklace} 
The collapse commences with the formation of nascent-clusters at sites with relatively higher densities along the chain. These clusters subsequently start coarsening by pulling 
monomers from the chains connecting them and eventually coalesce with each other, forming bigger clusters. This second stage of the collapse finally ends 
when all the monomers are in a single cluster. Strikingly one can observe in Fig.\ \ref{R_g}(b), that for Model\ II
the coalescence of clusters not only occurs along the chain, but also occurs from the lateral movement of the clusters. One can attribute this to the choice of much larger $N(=8192)$, 
which at intermediate times gives rise to structures that resembles branching or a network of ``pearls''.
Thus formation of multiple connections for a cluster to other clusters becomes quite feasible. 
Although this has not been observed in any previous simulation,\cite{kuznetsov1995kinetics,majumder2015cluster,majumder2017kinetics} courtesy of using smaller $N$, one must expect that 
in the thermodynamic limit ($N \rightarrow \infty$) this could be the real picture.
\par
Next we investigate the scaling of the relaxation time for both the models followed by the study of scaling of cluster coarsening. 
In the final subsection, we present results related to aging using a suitable two-time correlation function. 

\subsection{Relaxation Time}
Following the general trend used in most of the dynamic studies, we start our analyses with the understanding of the decay of the radius of gyration with time. 
The squared radius of gyration
\begin{equation}
 R_g^2=\frac{1}{2N^2}\sum_{i,j=1}^{N}(\boldsymbol{r}_i-\boldsymbol{r}_j)^2,
\end{equation} 
where $\boldsymbol{r}_i$ is the position of $i$-th monomer, is a measure of the spatial extension or size of a polymer. The plots in Figs.\ \ref{R_g}(a) and (b) show the decay of $R_g^2$ as a function of time $t$ for Model\ I 
and Model\ II, respectively. For off-lattice models both with \cite{polson2002simulation} and without hydrodynamics,\cite{majumder2017kinetics} it has 
been shown that such a decay of $R_g^2$ can be well described as 
\begin{equation}
R_g^2(t)=b_0+b_1\exp\left[ -\left(t/\tau_c\right)^\beta \right],
 \label{EqFit}
\end{equation}
where $b_0$ corresponds to the value of $R_g^2$ in the collapse phase, $b_1$ and $\beta$ are two non-trivial fitting parameters, and $\tau_c$ is a measure of the relaxation time or 
the collapse time. 
In Figs.\ \ref{R_g}(a) and (b), the solid lines are the corresponding best fits to \eqref{EqFit}, showing more or less a consistent behavior. 
The fits for all polymer lengths provided a reduced chi-squared value $\chi_r^2$ ($\chi^2$ per degree-of-freedom) close to unity (see Table \ref{TableFit}) suggesting \eqref{EqFit} as an appropriate analytic form.
For the estimation of the statistical error on the relaxation times $\tau_c$ from \eqref{EqFit} we performed a Jackknife analysis\cite{efron1982jackknife,efron1994introduction} to mitigate the effect of temporal correlations.
We note that the stretching exponent $\beta$ assumes a constant value of $\beta \approx 1.2$ in Model\ I,
while in Model\ II we observe a slight variation from $\beta \approx 1.0$ to $\beta \approx 1.15$ for increasing $N$, similar to the behavior observed in the off-lattice polymer model.\cite{majumder2017kinetics}
\begin{figure}[t!]
\includegraphics{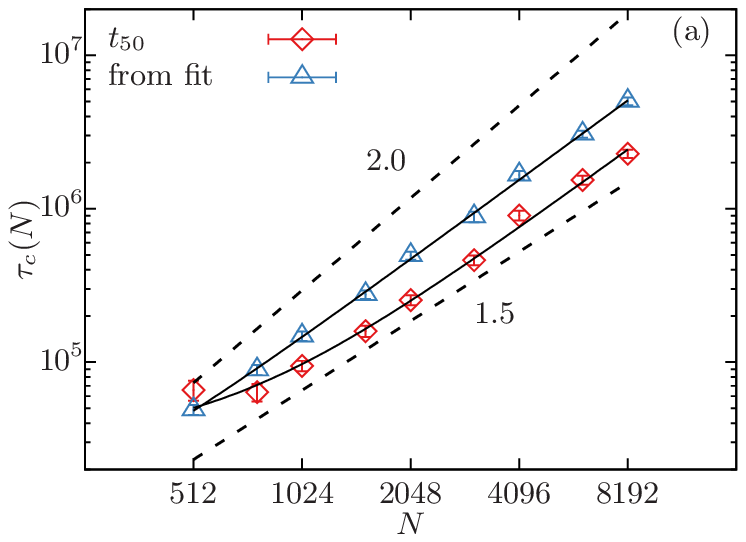}
\includegraphics{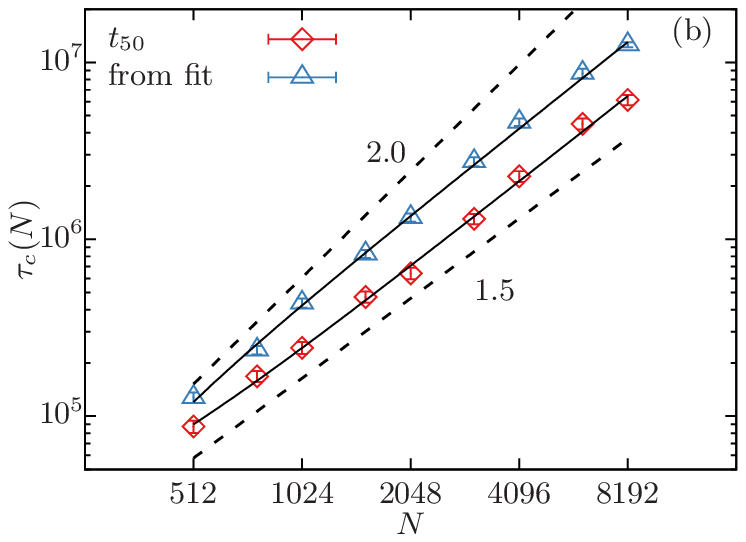}
\caption{Double-log plots showing the scaling of the relaxation time $\tau_c$ obtained from two different methods for (a) Model\ I and (b) Model\ II at $T_q=2.5$. 
The solid lines are the best fits of \eqref{tau_N}, with parameters as mentioned in Table\ \ref{TableExponent}. 
The dashed lines represent the Rouse behavior $\tau_c \sim N^2$ respectively the theoretical prediction $\tau_c \sim N^{1.5}$ of Ref.\ \onlinecite{Abrams2002}.
}
\label{Relaxation}
\end{figure} 
One can also estimate the collapse time $\tau_c$ by measuring the time $t_{50}$ when $R_g^2(t)$ has decayed to $\left[R(0)-R(\infty)\right]/2$, i.e., half of its total decay. 
The collapse times obtained from the fitting of Eq.\ \eqref{EqFit} and as $t_{50}$ are plotted in Figs.\ \ref{Relaxation}(a) and (b) for Model\ I and Model\ II, respectively. Apparently the data show 
a power-law behavior which one can quantify by using the form 
\begin{eqnarray}\label{tau_N}
\tau_c=B N^z+\tau_0,
\end{eqnarray}
where $B$ is a nontrivial constant which may depend on the quench temperature $T_q$, 
$z$ is the corresponding dynamic critical exponent, and the offset $\tau_0$ comes from finite-size corrections. 
In absence of hydrodynamics, the Rouse model \cite{rouse1953theory} predicts $z=2$ for such relaxation in equilibrium dynamics. 
In previous studies of such dynamic exponent in equilibrium dynamics of a lattice polymer with relatively smaller $N$ a value of $z\approx 2.1$ was reported.\cite{gurler1983effect} 
The results of fitting the two different collapse times for both models to the form \eqref{tau_N} are tabulated in Table\ \ref{TableExponent}. 
For Model\ I (Model\ II) fitting of the data provides $z \approx 1.73$ ($z \approx 1.62$). 
These values are somewhat larger but still compatible with the value ($z=1.5$) predicted in a theory using a coarse-grained picture of the collapsing polymer in absence of hydrodynamics.\cite{Abrams2002} 
Later, we show a possible algebraic connection of $z$ with the cluster-growth exponent $\alpha_c$.

\begin{table}[t!]
\caption{Results obtained from the Jackknife analysis of fits of the form \eqref{EqFit} to the decay of the squared radius of gyration $R_g^2(t)$, for Model\ I and Model\ 
II using three different polymer lengths. The reduced chi-squared $\chi_r^2$ is the average goodness-of-fit parameter.}
\centering
\begin{tabular}{ccccc}
\hline
\hline
$N$ & Model  & $\tau_c$ ($10^5$ MCS) & $\beta$ & $\chi_r^2$ \\
\hline
$2048$ & I & $5.0(3)$ & $1.23(5)$ & $1.5(2)$\\
$4096$ & I & $16(1)$ & $1.22(5)$  &$1.5(2)$ \\
$8192$ & I & $50(3)$ & $1.23(4)$ & $1.7(3)$\\
\hline
$2048$ & II & $13.3(8)$ & $1.00(5)$ & $0.5(3)$\\
$4096$ & II & $46(3)$ & $1.15(5)$ & $1.3(7)$\\
$8192$ & II & $126(4)$ & $1.14(4)$ & $2.0(2)$\\
\hline
\hline
\end{tabular}
\label{TableFit}
\end{table}

\begin{table}[t!]
\caption{Results obtained from the fits of \eqref{tau_N} to the relaxation times $\tau_c$ for the full range of data $N \in [512,8192]$, using both the models. 
Here again $\chi_r^2$ measures the goodness of fit.}
\centering
\begin{tabular}{ccccc}
\hline
\hline
Method & Model & $z$ & $\chi_r^2$ \\
\hline
$t_{50}$ & I & $1.72(8)$ & $1.61$\\
from fit & I & $1.73(5)$ & $0.56$\\
\hline
$t_{50}$ & II & $1.61(5)$ & $1.08$\\
from fit & II & $1.62(4)$ & $1.50$\\
\hline
\hline
\end{tabular}
\label{TableExponent}
\end{table}

\subsection{Cluster-Growth Kinetics}
Now, after having an idea on the relaxation times related to the collapse, we shift our focus to the cluster-growth kinetics. 
The formation and growth of clusters of monomers bear certain resemblance with the coarsening of particle or spin systems.\cite{Bray2002,Majumder_Ising2010,Majumder_Ising2011,puri2004kinetics} 
As already mentioned, this fact has been exploited to understand the collapse dynamics in an off-lattice model polymer.\cite{majumder2015cluster,majumder2017kinetics} Following this approach, for the 
lattice models considered here, the ordered structures (clusters of monomers) can be detected. Thus the formation and growth of clusters can be monitored 
by measuring the number of clusters $N_c$ and the corresponding size of clusters $C_s(t)$ as a function of time during the collapse. 
The identification of clusters is achieved by iterating over the polymer and identifying the number of monomers $n_i$ within a distance $r_{\text{max}}$ around each monomer $i$. 
If this number of monomers $n_i$ exceeds a certain minimum number of monomers ($n_i>n_{\text{min}}$), then there is said to be a cluster around monomer $i$ containing all the $n_i$ monomers. 
The overlap of clusters (a single monomer cannot be part of two or more clusters) introduced by this method is resolved by assigning overlapping clusters into a single cluster. 
Different combinations of $n_{\text{min}}$ and $r_{\text{max}}$ produces comparable results in the coarsening regime of cluster growth for reasonable choices. For the results to be
presented here we opt for $n_{\text{min}}=10$ and $r_{\text{max}}=2$ as in Ref.\ \onlinecite{kuznetsov1995kinetics}. Moreover, with the aim of exploiting the advantages of a lattice model to 
identify ordered structures as well as to characterize the morphology, we calculate a two-point equal-time density-density correlation function, defined as
\begin{equation}\label{corrfunc}
C(r,t)=\langle \rho_i(0,t) \rho_i(r,t) \rangle,
\end{equation} 
where 
\begin{equation} 
\rho_i(r,t)=\frac{1}{m_r}\sum_{j, r_{ij}=r} \theta(\boldsymbol{r}_j,t).
\end{equation} 
The characteristic function $\theta$ is unity if there is a monomer at a position $\boldsymbol{r}_j$ or zero otherwise.
The number of possible lattice points at distance $r$ from an arbitrary point of the lattice is denoted by $m_r$, which obviously is dependent 
on the type of underlying lattice.

\begin{figure}[t!]
\includegraphics{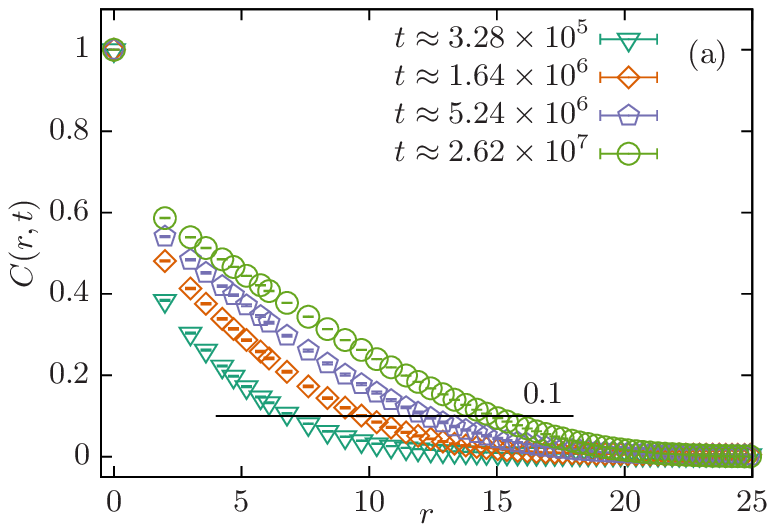}
\includegraphics{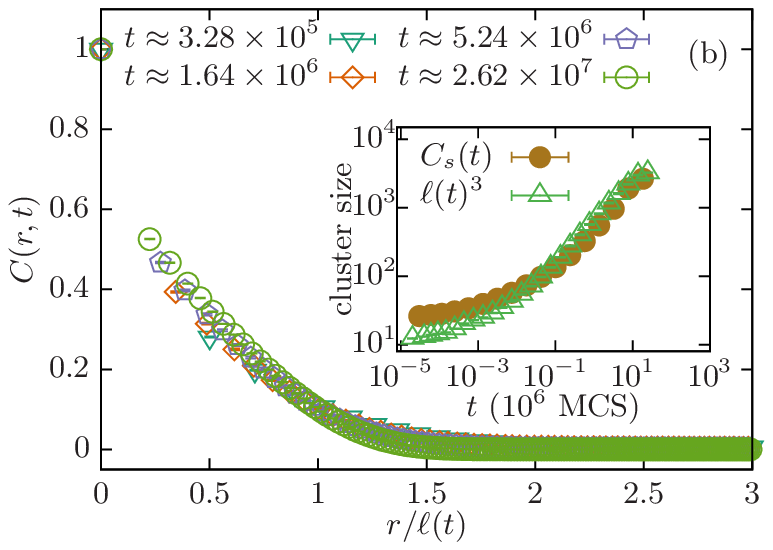}
 \caption{(a) Two-point equal-time correlation function $C(r,t)$, at different times for Model\ II. The solid line depicts the extraction of the characteristic length 
 $\ell(t)$ using $h=0.1$ in \eqref{lenght_from_corr}. (b) The corresponding scaling plots as a function of $r/\ell(t)$. The inset of (b) shows the comparison of $\ell(t)^3$ and the cluster size $C_s(t)$ on a log-log scale. 
 All the results are obtained for $N=4096$ monomers for a quench at $T_q=2.5$. The time $t$ is denoted in units of Monte Carlo sweeps (MCS).}
 \label{CorrelationFig}
\end{figure}
In Fig.\ \ref{CorrelationFig}(a) we plot such exemplary $C(r,t)$ at four different times for Model\ II quenched to $T_q=2.5$. The signature of a presence of a growing 
nonequilibrium length scale is clearly evident as the curves decay farther with the increase of time. For Model\ I and different quench temperatures $T_q$ we observe similar behavior, however, we do not 
present such plots for the sake of brevity. Following the exercise prevailed in phase-ordering business,\cite{Bray2002,puri2004kinetics} we extract 
an average length scale $\ell(t)$ that describes the ordering, i.e., clustering during the collapse, using the criterion 
\begin{equation}\label{lenght_from_corr}
C\left[r=\ell(t),t \right]=h,
\end{equation} 
where we chose $h=0.1$. Other values of $h$ produce a proportional behavior.
For the characteristic length $\ell(t)$, following the trend in ordering phenomena studies, one 
looks for the scaling of the form
\begin{equation}
C(r,t) \equiv \tilde{C}(r/\ell(t)),
\label{EqEquiv}
\end{equation}
where $\tilde{C}$ is the scaling function.
The presence of such scaling is demonstrated in Fig.\ \ref{CorrelationFig}(b), showing the collapse of $C(r,t)$ for Model\ II at different times when 
plotted as function of $r/\ell(t)$. Although we do not present it here, Model\ I shows a similarly good scaling.
\par

Since the ordering during collapse is manifested by formation of the clusters of monomers, the characteristic length $\ell(t)$ must be related to the cluster size $C_s(t)$, 
obtained via the cluster recognition method such that
\begin{equation}
C_s(t) \propto \ell(t)^{d_f},
\label{EqClusterGrowth} 
\end{equation}
where $d_f$ is the fractal dimension of the clusters. To estimate this fractal dimension, we set the radius of gyration for each single cluster in relation to its mass 
(number of monomers) and obtain $d_f \approx 3$. In the inset of Fig.\ \ref{CorrelationFig}(b) we compare $\ell(t)^3$ with $C_s(t)$ for 
Model\ II at quench temperature $T_q=2.5$. Apart from a little discrepancy at early times they seem to be proportional to each other. Based on this observation, 
hence, from now onward, we will use $\ell(t)^3$ to characterize the cluster growth.
In particular this has advantages at comparatively higher quench temperatures $T_q$, 
where the cluster identification method fails to recognize the final globular structure as a single cluster for a single chosen set of $n_{\text{min}}$ and $r_{\text{max}}$.
\par
\begin{figure} 
\includegraphics{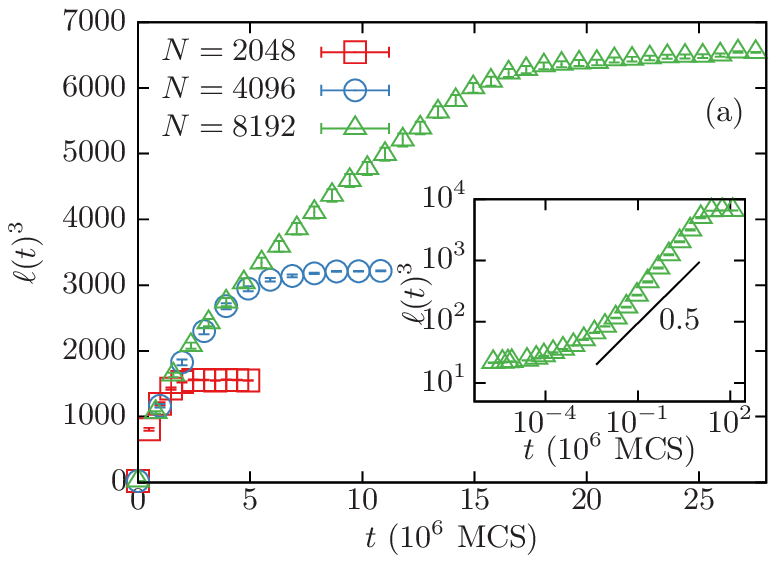}
\includegraphics{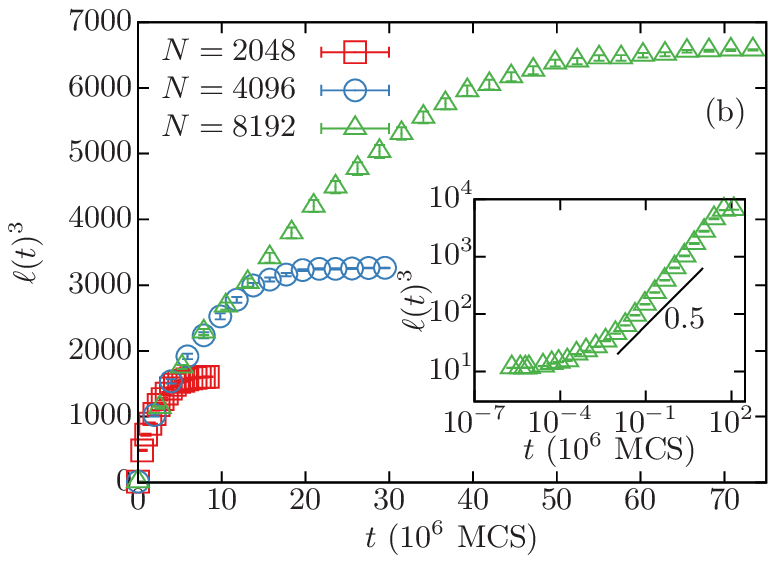}
\caption{Plot of the cubed characteristic length $\ell(t)^3$, as a measure of the average cluster size, against time for (a) Model\ I and 
(b) Model\ II with $N=2048,4096,$ and $8192$ quenched to $T_q=2.5$. The insets show the data for $N=8192$ on a double-log scale. The solid lines there show power-law behaviors with exponent $1/2$.}
\label{GrowthCharact}
\end{figure}

Now, from the scaling of $C(r,t)$ one can easily deduce the fact that the ordering or rather the cluster growth follows a power-law scaling $\ell(t)^3 \sim t^{\alpha_c}$. 
As already mentioned,\cite{majumder2015cluster,majumder2017kinetics} due to the involvement of crossover from the initial cluster formation stage, the true scaling behavior 
is described by \eqref{cluster_growth}. We start our quantification by replacing $C_s(t)$ with $\ell(t)^3$, where $\ell(t)$ is extracted using \eqref{lenght_from_corr}, in \eqref{cluster_growth} 
to obtain 
\begin{equation}\label{length_cluster}
 \ell(t)^3=\ell_0^3+At^{\alpha_c}. 
\end{equation}
Figure\ \ref{GrowthCharact} shows the time dependence of $\ell(t)^3$ for different polymer lengths, as indicated, 
with the same data shown for only $N=8192$ on a double-log scale in the inset, for (a) Model\ I and (b) Model\ II. 
The data for different $N$ follow each other until finite-size effects become apparent. 
The finite-size effect creeps in when all the monomers become part of a single cluster, and thereby no further growth is observed.
Initially there is a transition period in the growth (can be seen from the double-log scale plot), 
marking the initial stage of cluster formation. After the initial clusters are formed, the growth crosses over to the coarsening regime 
where it indeed shows a power-law scaling. The data in both cases seem to have a higher slope than the solid lines with exponent $1/2$, as observed in Ref.\ \onlinecite{kuznetsov1995kinetics}. 
We use the form \eqref{length_cluster} to fit the data, considering the crossover in the growth. 
This yields \begin{equation*}\alpha_c=0.68(6)\text{ for Model\ I}\end{equation*}and \begin{equation*}\alpha_c=0.62(5)\text{ for Model\ II.}\end{equation*} 

\subsubsection{Temperature-Dependence of the Cluster Growth}
\begin{figure}[t!]
\includegraphics{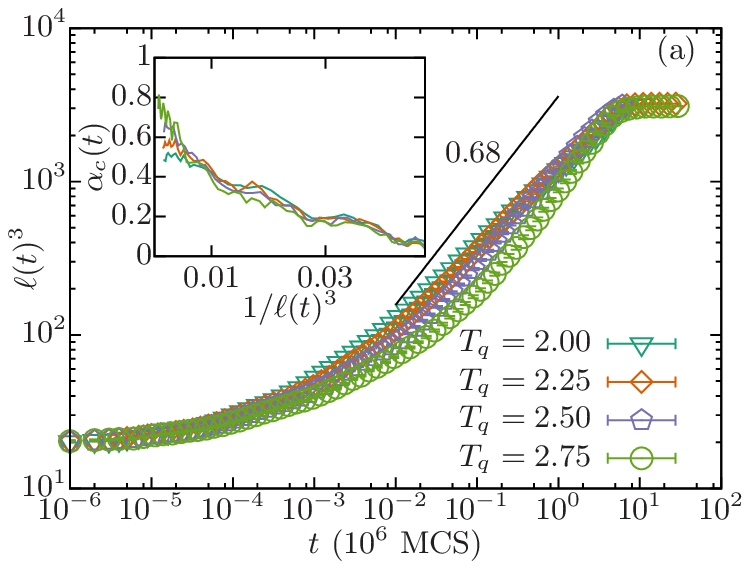}
\includegraphics{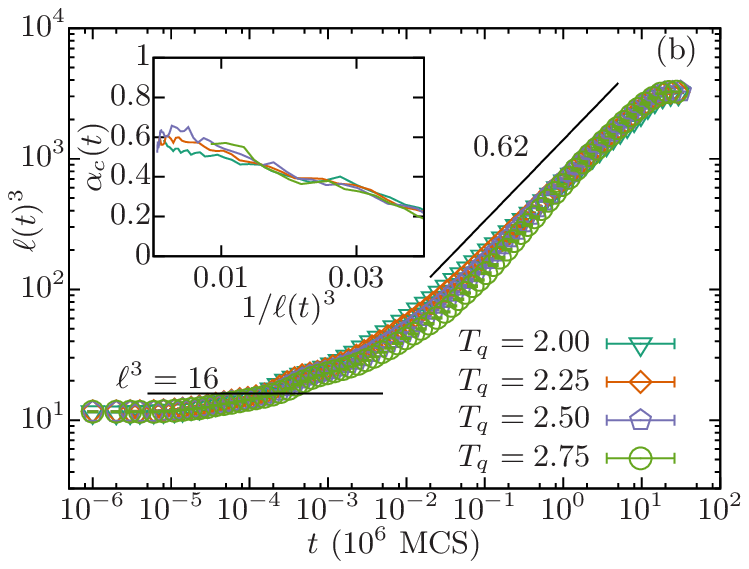}
\caption{Double-log plot showing the dependence of the cluster growth on the quench temperature $T_q$ in (a) Model\ I and (b) Model\ II, measured as $\ell(t)^3$ for $N=4096$. 
The inset shows the respective time-dependent instantaneous exponent $\alpha_c(t)$ calculated as described in Eq.\ \eqref{ins_expo}.}
\label{TempMod1}
\end{figure}
Next we turn to a check of the robustness of the kinetics for both the models, viz., the influence of the quench temperature $T_q$.
In simulations of various off-lattice models it has been shown, that changing temperature correctly reproduces the change 
in solvent quality, with faster collapse as the temperature decreases.\cite{polson2002simulation,majumder2017kinetics} In contrast, here we find that a decrease in 
temperature results in a slower collapse, similar to the behavior observed for ordering dynamics of the Ising model,\cite{majumder2013temperature} where 
faster equilibration occurs at higher temperature due to increasing diffusion of particles. In addition, for low enough temperatures the system may get 
trapped in some metastable state with high energy barrier, quite difficult to overcome via simple Metropolis dynamics. Thus to avoid such situations, 
we restrict ourselves to simulations at relatively higher temperatures $T_q \in [2.0,2.75] $.
\par
We show the growth of clusters for Model\ I at different $T_q$ in the main frame of Fig.\ \ref{TempMod1}(a). It is apparent that, within the chosen range 
of temperature, the scaling of the growth is non-universal in nature. On the other hand, for Model\ II, shown in the main frame of Fig.\ \ref{TempMod1}(b) the growth looks quite independent of $T_q$. 
A fitting with the form \eqref{length_cluster} provides $\alpha_c$ having a wide 
range $[0.5,0.8]$ for Model\ I whereas for Model\ II a much narrower range $[0.58,0.65]$ is obtained. In this regard, we also calculate the instantaneous exponent $\alpha_c(t)$ given as 
\begin{eqnarray}\label{ins_expo}
\alpha_c(t)=\frac{d\ln \ell(t)^3}{d \ln t}, 
\end{eqnarray}
which when operated on \eqref{length_cluster} yields 
\begin{eqnarray}\label{asymptotic_alpha}
\alpha_c(t)=\alpha_c \left[1-\frac{\ell_0^3}{\ell(t)^3} \right]. 
\end{eqnarray}
Thus a plot of $\alpha_c(t)$ against $1/\ell(t)^3$ would provide the asymptotic $\alpha_c$ in the limit $1/\ell(t)^3\rightarrow 0$. 
In the insets of Figs.\ \ref{TempMod1}(a) and (b) we show such plots of $\alpha_c(t)$ for the respective models. The asymptotic 
behavior clearly indicates that the growth in Model\ II is of more universal nature. Like in Model\ I, temperature-dependent growth exponents were 
earlier observed in quenches of an Edwards-Anderson spin glass in $d=3$,\cite{komori1999numerical} in an anisotropic rotor model with
vacancies \cite{mouritsen1985temperature} as well as in disordered ferromagnets.\cite{paul2004domain,paul_PRE2005,Paul2007} 
This connection might be reasonable, as all the concerned systems are dominated by disorder and constraints of the lattice structure.
On the other hand, perhaps the introduction of bond fluctuation due to consideration of the diagonal bonds helps to overcome 
the topological constraints of the lattice to some extent, hence, a temperature-independent growth at moderately high temperatures.
\par
For both models the growth exponent is smaller than in the off-lattice model where $\alpha_c \approx 1$. \cite{majumder2015cluster,majumder2017kinetics}
For Model\ I, however, the value of $\alpha_c$ is not universal, as the growth exponent appears to be dependent on the quench temperature $T_q$.
The value obtained for Model\ II already gives an indication for a different (temperature-independent) growth exponent than $1/2$ as reported in Ref.\ \onlinecite{kuznetsov1995kinetics}. 
For further investigation we call for a nonequilibrium finite-size scaling analysis for Model\ II.

\subsubsection{Finite-Size Scaling Analysis}
Using a finite-size scaling analysis one aims at extracting quantities in the thermodynamic limit ($N \rightarrow \infty$) from simulations of a finite size. 
In simulations we necessarily have finite systems and thus this analysis method has its wide application in the context of critical phenomena.\cite{Fisher_book}
Later finite-size scaling has also been successfully adapted to the nonequilibrium scenario to understand the growth exponents in particle \cite{das2012finite} and 
spin \cite{Majumder_Ising2010,Majumder_Ising2011} systems. Here, we rely on such an exercise previously performed in more detail for the off-lattice model in Ref.\ \onlinecite{majumder2017kinetics}. This method was applied successfully
to understand the collapse dynamics. The finite-size scaling ansatz is constructed by expanding Eq.\ \eqref{length_cluster} to additionally include an initial crossover time $t_0$,
\begin{equation}\label{FS_ansatz}
\ell(t)^3=\ell_0^3+A(t-t_0)^{\alpha_c}.
\end{equation} 
The values of $\ell_0$ and $t_0$ are marking the point after which the coarsening regime starts, and are analogous to the 
background contribution in critical phenomena. Following Ref.\ \onlinecite{majumder2017kinetics}, we identify the linear cluster size 
$\ell(t)$ ($\sim C_s(t)^{1/3}$) with the equilibrium correlation length $\xi$, and $1/t$ with the temperature deviation from a critical point. 
In order to account for the finite-size effect in \eqref{FS_ansatz}, one can write down \cite{majumder2017kinetics}
\begin{equation}
\ell(t)^3-\ell_0^3=(\ell_{\max}-\ell_0)Y(y)
 \label{FS_eqn}
\end{equation} 
where the finite-size scaling function 
\begin{equation}
 Y(y)=\frac{\ell(t)^3-\ell_0^3}{\ell_{\max}^3-\ell_0^3}
 \label{EqScaling1}
\end{equation} 
and the scaling variable 
\begin{equation}
y=\frac{(\ell_{\max}^3-\ell_0^3)^{\frac{1}{\alpha_c}}}{t-t_0}.
\label{EqScaling2}
\end{equation} 
Note that in \eqref{FS_eqn}, \eqref{EqScaling1} and \eqref{EqScaling2} $\ell_{\max}$ is the saturation 
value of $\ell(t)$, which one obtains when all the monomers of the polymer are in a single dense globular cluster, and does not need to be equal to $N^{1/3}$, 
but rather proportional, $\ell_{\max} \sim N^{1/3}$.
In the finite-size unaffected regime, the form \eqref{FS_ansatz} is recovered, providing 
\begin{equation}\label{master_curve}
 Y(y) \propto y^{-\alpha_c},
\end{equation} 
while in the finite-size affected regime one must obtain $Y(y) \rightarrow 1$. 
\begin{figure}
\includegraphics{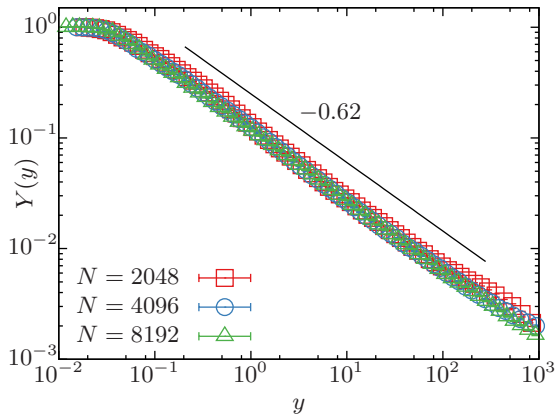}
\caption{Finite-size scaling plot for Model\ II at $T_q=2.5$ using the ansatz \eqref{FS_ansatz}. The solid line corresponds to the expected power-law decay with an 
exponent $\alpha_c=0.62$.}
\label{FSS_model2}
\end{figure}
In the finite-size scaling exercise we tune the value of $\alpha_c$ to obtain an optimum collapse of data for different $N$, obeying
the master curve behavior \eqref{master_curve}. From the fitting exercise done in the previous section we have a fair idea about $\ell_0$. 
We chose $\ell_0^3 \approx 16$ and $t_0 \approx 200$ for Model\ II, independent of $N$. The exercise 
yields a reasonably good collapse of data and corresponding master curve behavior with 
$\alpha_c=0.62(3)$ for Model\ II, in agreement with the direct fit. In Fig.\ \ref{FSS_model2} 
we present a representative plot for Model\ II. The corresponding plot for Model\ I is omitted due to the quench-temperature dependency.
On one hand the value obtained is different from the previously reported value of $1/2$ for a lattice polymer.
\cite{kuznetsov1995kinetics} On the other hand $\alpha_c\approx 2/3$ is comparable with the value
predicted for an off-lattice model via Gaussian self-consistent theory,\cite{kuznetsov1996kinetic} confirmed by Langevin dynamic simulations.\cite{byrne1995kinetics}
On the contrary, for a similar off-lattice model, however, one observes a linear cluster growth \cite{majumder2015cluster,majumder2017kinetics} 
as observed for Ostwald ripening (corresponding to the Ostwald exponent $1/3$ when considering the length scale). 
Now, by using the fact that at the point of onset of finite-size effects $\ell(t)^3 \sim N$ and by replacing the 
corresponding time $t$ as the collapse time $\tau_c$, one can show \cite{majumder2017kinetics} from \eqref{EqScaling1} and \eqref{EqScaling2} that $z=1/\alpha_c$. 
This relation holds quite nicely for Model\ II as it yields  $z \approx 1.61$, consistent with the behavior shown in Fig.\ \ref{Relaxation} and Table\ \ref{TableExponent}.

\begin{figure}[t!]
 \includegraphics{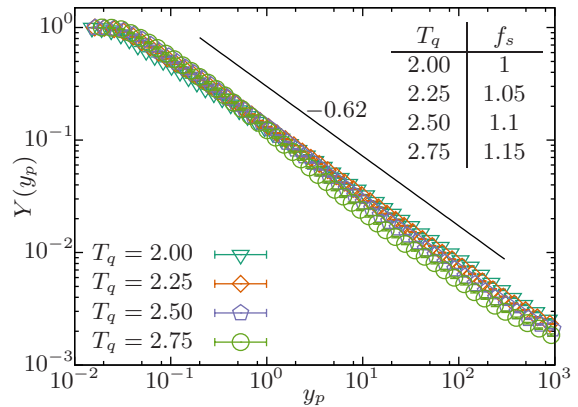}
 \caption{Scaling plot for Model\ II for different quench temperatures $T_q$, where $f_s$ is the metric factor defined in Eq.\ \eqref{metric_factor}. 
 The solid line corresponds to a power-law function with exponent $\alpha_c=0.62$.}
 \label{TempMod2}
\end{figure}

\subsubsection{Temperature-Dependent Scaling of the Cluster Growth}\label{TempGrowth}
To further look into the universal nature of the scaling in Model\ II, we apply a modified scaling analysis based 
on the above discussed finite-size scaling analysis. Here we account for the different growth amplitudes by modifying the scaling variable \eqref{EqScaling2} as \cite{majumder2017kinetics}
\begin{equation}
 y_p=f_s\frac{(\ell_{\max}^3-\ell_0^3)^{\frac{1}{\alpha_c}}}{t-t_0},
\end{equation}
with a metric factor depending on the growth amplitudes
\begin{equation}\label{metric_factor}
f_s(T_q)=\left(\frac{A(T_q=2.0)}{A(T_q)}\right)^{\frac{1}{\alpha_c}}.
\end{equation}
Note that $y_p$ differs from $y$ only by this factor $f_s$. A fitting of the $\alpha_c(t)$ presented in the inset of Fig.\ \ref{TempMod1}(b) to the scaling law \eqref{asymptotic_alpha} provides 
a rough estimate of $\ell_0^3=16 \pm 2$ for all quench temperatures, consistent with the previously mentioned value of $\ell_0^3 \approx 16$ for the polymer quenched to $T_q=2.5$.
We use this value of $\ell_0^3$ and the corresponding $t_0$ values in the scaling exercise and tune the value 
of $\alpha_c$ such that the data for different $T_q$ collapse onto a single master curve for appropriate adjustments of the metric factor $f_s$, i.e., the ratio of amplitudes. 
Recall that an appropriate choice of $\alpha_c$ should lead to a consistent power-law behavior of the finite-size scaling function as $Y(y_p) \sim y_p^{-\alpha_c}$ along 
with optimum data collapse. In our exercise we obtained such behavior for $\alpha_c=0.62(4)$. In Fig.\ \ref{TempMod2}, we show such a representative plot for $\alpha_c=0.62$. 
The successful application of such a scaling exercise thus indicates that indeed the scaling of cluster growth in the Model\ II is nearly universal in nature 
which can be described by a universal finite-size scaling function with a nonuniversal metric factor. This observation has recently been made in an off-lattice model,\cite{majumder2017kinetics,majumder2015cluster} however, 
for a linear scaling of the cluster growth.

\subsection{Aging and Related Scaling}
 \begin{figure}
  \includegraphics{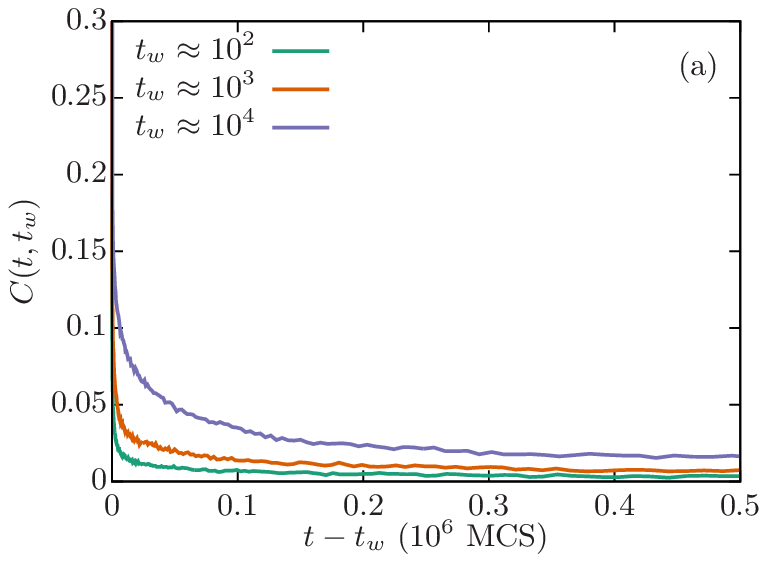}
  \includegraphics{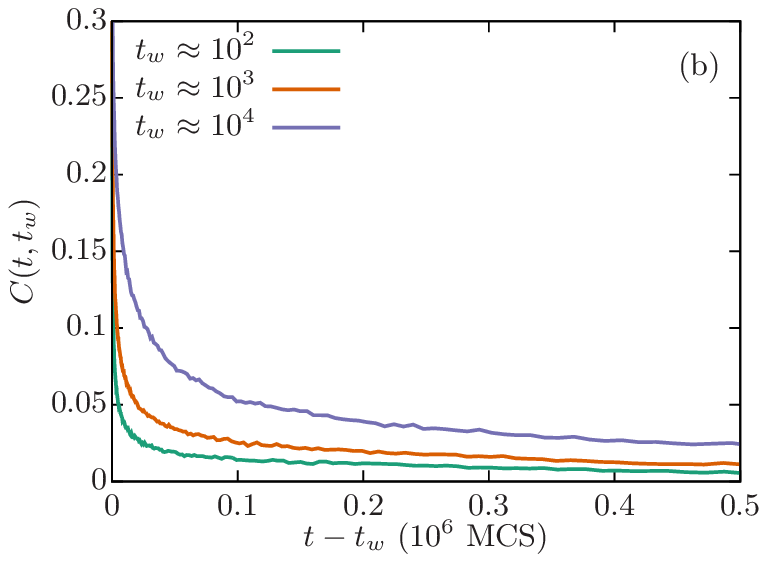}
  \caption{Plot of two-time correlation functions $C(t,t_w)$ against $t-t_w$ for (a) Model\ I and for (b) Model\ II. 
  The exact values of $t_w$ are very close to the indicated value. The length of the polymer used here is $N=8192$ and the corresponding quench temperature is $T_q=1.5$.}
  \label{Figttw}
 \end{figure}

\begin{figure*}
  \includegraphics{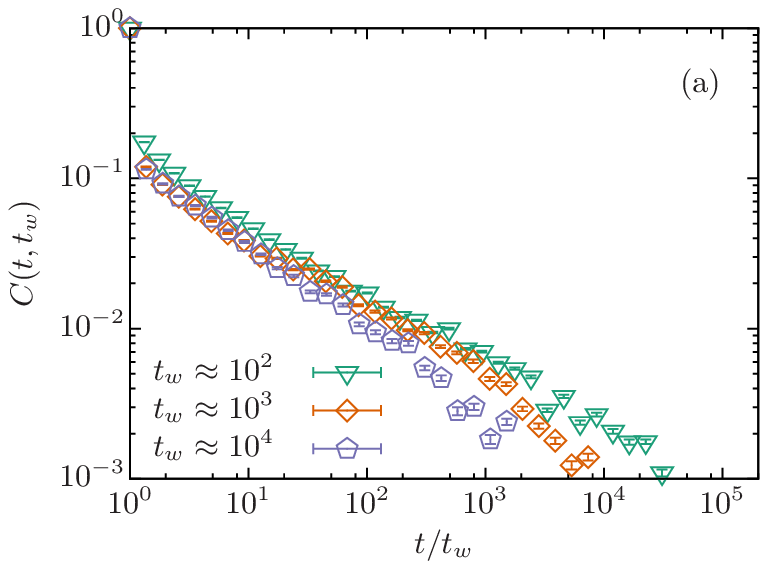}\includegraphics{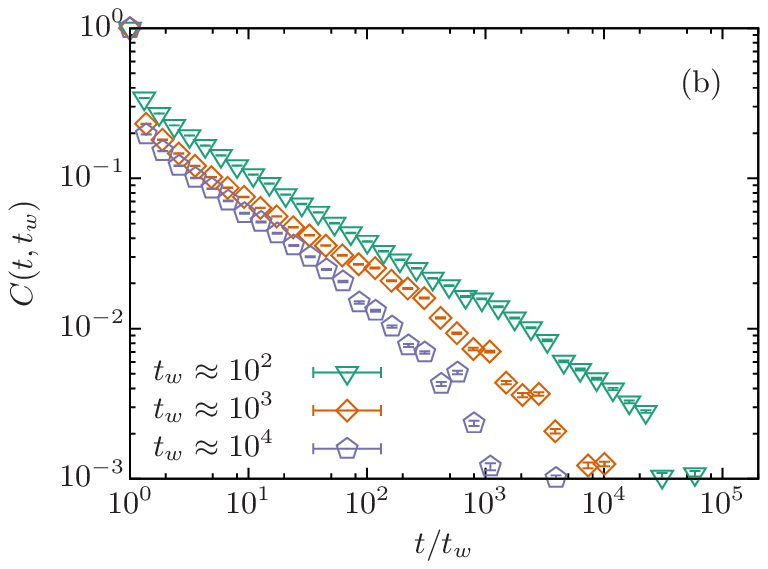}
  \includegraphics{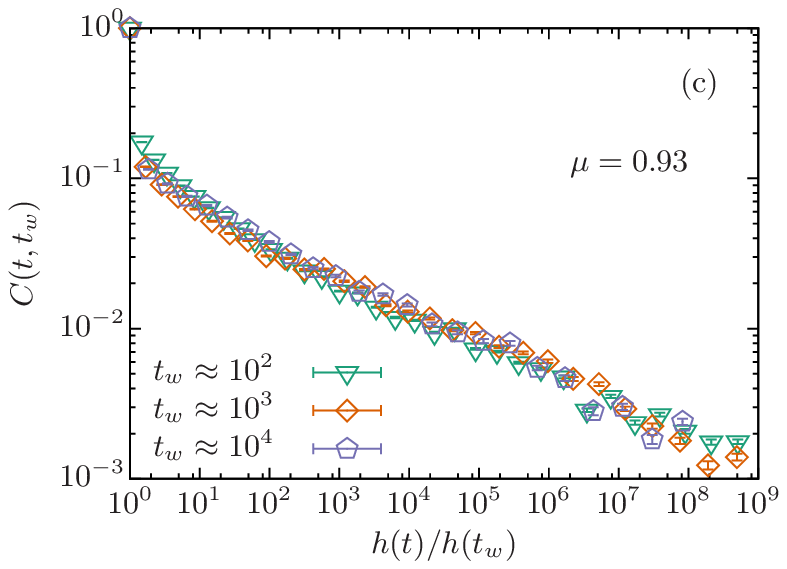}\includegraphics{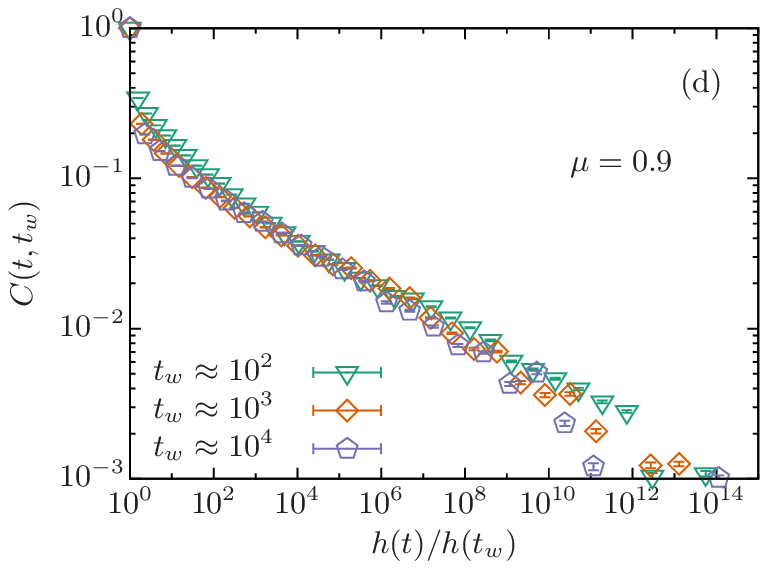}
  \includegraphics{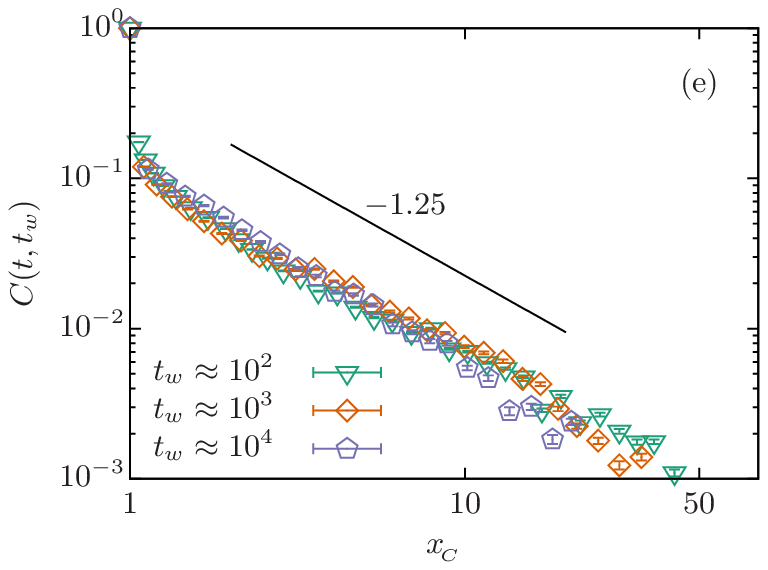}\includegraphics{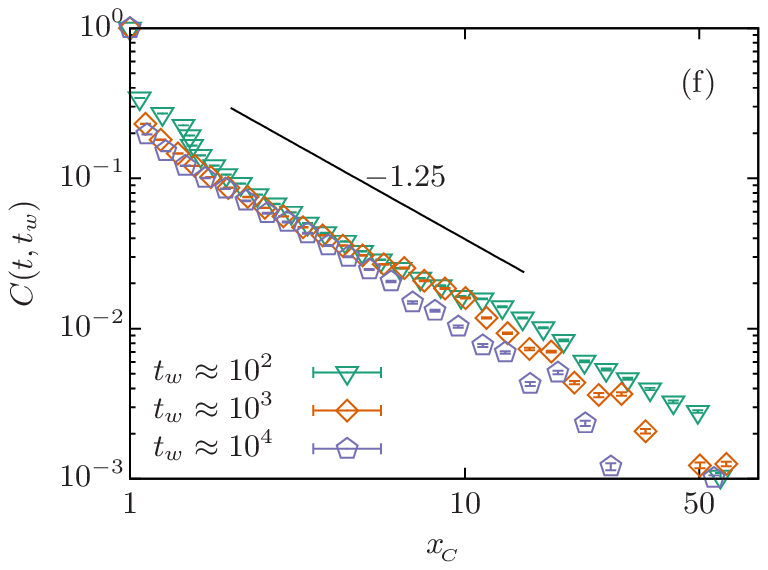} 
  \caption{The scaling of the two-time correlation function $C(t,t_w)$ with respect to $t/t_w$ is shown in (a) for Model I and (b) for Model II. The scaling 
  of form \eqref{superaging} is shown in (c) for Model\ I and (d) for Model\ II. The value of $\mu$ was obtained by trial and error to get the best data collapse.
  In (e) for Model\ I and (f) for Model\ II the scaling with respect to the ratio of the cluster sizes, i.e, $x_{\!_C}=\ell(t)^3/\ell(t_w)^3$ is presented. 
  The solid lines in those plots show a power-law decay with slope $-1.25$. 
  In all plots of this figure the same data is presented, as can be seen from the range of values for the two-time correlation function $C(t,t_w)$.
  The length of the polymer used here is $N=8192$ and the corresponding quench temperature is $T_q=1.5$.}
  \label{Figtbytw}
 \end{figure*}

Until now, our focus has been solely on equal-time quantities governing the kinetics. Here, in this subsection we 
turn our attention to the behavior of a two-time quantity, used to probe aging in 
an evolving nonequilibrium system. Using the framework recently developed for an off-lattice model,\cite{majumder2016evidence,majumder2016jpcm,majumder2017kinetics} 
we construct the two-time correlation function as
 \begin{equation}
 C(t,t_w)=\langle O_i(t)O_i(t_w) \rangle-\langle O_i(t) \rangle \langle O_i(t_w) \rangle.
 \end{equation}
 We assign $O_i=\pm 1$ by checking the radius $r$, at which the local density, given by $\rho_i(r,t)$ (see Eq.\ \eqref{corrfunc}), first falls below a threshold of $0.1$. 
 If this radius is smaller than $\sqrt{3}$ we assign $O_i=1$, marking a high local density, otherwise we chose $O_i=-1$ to mark a low local density.
 This definition is an adaptation of the method used in Refs.\ \onlinecite{majumder2016evidence,majumder2017kinetics}, where one relies on the cluster identification method.
 Nonetheless, both methods are analogous to the usual two-time density-density correlation function in particle systems.
 In Fig.\ \ref{Figttw}(a), $C(t,t_w)$ is plotted against the translated time $t-t_w$ at different values of the waiting time $t_w$ for Model\ I and in (b) for Model\ II 
 with a polymer of length $N=8192$, quenched to $T_q=1.5$. Note here that the quench temperature $T_q$ is lower than in the previous section. 
 It will become clear as we move forward, that here the scaling is independent of $T_q$ for both the models.
 One can clearly see that the data for different $t_w$ do not overlap and the decay becomes slower with increasing waiting time $t_w$. 
 Such absence of time-translation invariance is a necessary condition for aging in the system.
In case of simple aging as described by \eqref{aging_scaling}, assuming an algebraic growth of the relevant length scale, one expects 
\begin{equation}\label{simpleaging_tw}
 C(t,t_w) \equiv g(t/t_w),
\end{equation} 
where $g$ is the scaling function of the variable $t/t_w$. 
While plotting $C(t,t_w)$ as a function of $t/t_w$ we fail to observe any scaling for both the models as shown in Figs.\ \ref{Figtbytw}(a) for Model\ I and (b) for Model\ II.

Such observation of ``no data collapse'' is known in ordering kinetics of diluted ferromagnets.\cite{Paul2007,park2010aging} This led the authors of 
Ref.\ \onlinecite{Paul2007} to use a special fitting ansatz given as 
\begin{equation}\label{superaging}
 C(t,t_w) \equiv G\left(\frac{h(t)}{h(t_w)}\right),
\end{equation} 
with the argument
\begin{equation}\label{superaging2}
h(t)=\exp\left(\frac{t^{1-\mu}-1}{1-\mu}\right).
\end{equation}
Here, $G$ is the scaling function and $\mu$ is a nontrivial exponent. Such an exercise with $\mu > 1$ indeed provided them an apparently reasonable collapse of the data. 
This is referred to as the so-called superaging behavior. 
However, it is known that algebraic constraints on the form of the autocorrelation function rule out the existence of such superaging behavior.\cite{kurchan2002elementary}
Later, while the analysis of Ref. \onlinecite{park2010aging} numerically convincingly suggested that such an ansatz may indeed be a 
good fitting function that provides a reasonable collapse of data, the true scaling behavior has been shown by these authors to be realized when one instead uses 
the generic scaling form \eqref{aging_scaling}.

 Observation of data collapse using \eqref{superaging} with $0 <\mu < 1$ is referred to as subaging, 
 observed mostly in soft matter systems.\cite{Ramos2001} Previously this has also been observed for a collapsing polymer.\cite{pitard2001glassy,Stanley2002}  
 We have tested the scaling ansatz of Eqs.\ \eqref{superaging} and \eqref{superaging2} with our data and tuned the exponent $\mu$ by trial and error until best data collapse is realized.
 We find $\mu \approx 0.9$ for both models. The corresponding scaling plots are shown in Figs.\ \ref{Figtbytw}(c) for Model\ I and (d) for Model\ II. 
 However, when we now simply plot $C(t,t_w)$ as a function of the ratio of cluster sizes $x_{\!_C}=\ell(t)^3/\ell(t_w)^3$, as shown in 
 Figs.\ \ref{Figtbytw}(e) and (f) for both the models, we also observe a reasonable collapse of data.\footnote{More generally one uses the scaling form $C(t,t_w) = C_s(t_w)^{-bz} f( C_s(t)/C_s(t_w))$,
where the non-trivial exponent $b$ is a priori unknown. For our system, however, we find $b=0$.} This implies that, in the present 
 case, too, the generic form \eqref{aging_scaling} describes the true scaling behavior rather than any special aging.
 
 In Ref.\ \onlinecite{park2010aging}, it has been argued that the simple aging of the form \eqref{simpleaging_tw} has been deduced 
 considering an algebraic growth of the relevant length scale. However, in their case a crossover from the algebraic growth to a slower logarithmic growth in the asymptotic limit makes 
 such deduction meaningless, hence, absence of simple scaling with respect to $t/t_w$. In the present case, 
 for clustering during the polymer collapse also, we encountered a crossover from a transient period of growth in the 
 initial cluster formation stage to a faster growth in the coarsening or coalescence stage, a fact that can be appreciated 
 from the insets shown in Fig.\ \ref{GrowthCharact}. 
 This reversal of crossover in our case could be the reason for the apparent appearance of subaging behavior with $\mu<1$, in contrast to the superaging behavior, where the crossover occurs the other way around.
 \begin{figure}[t!]
  \includegraphics{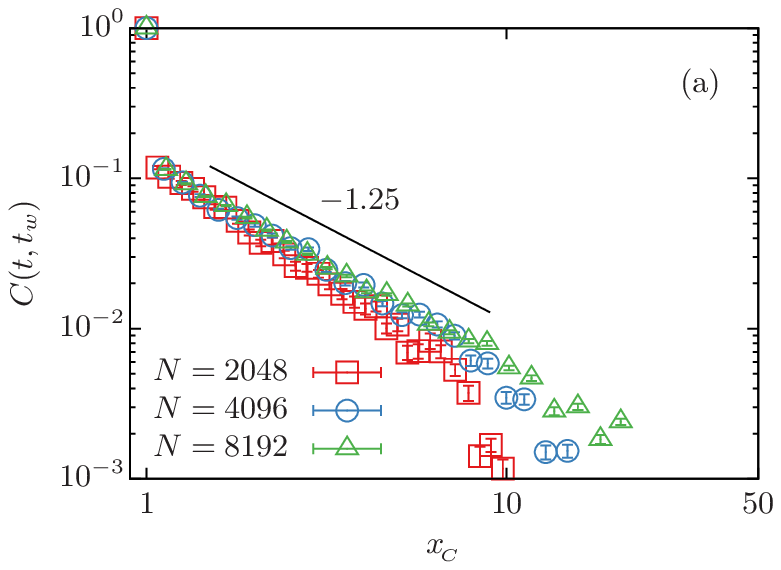}
  \includegraphics{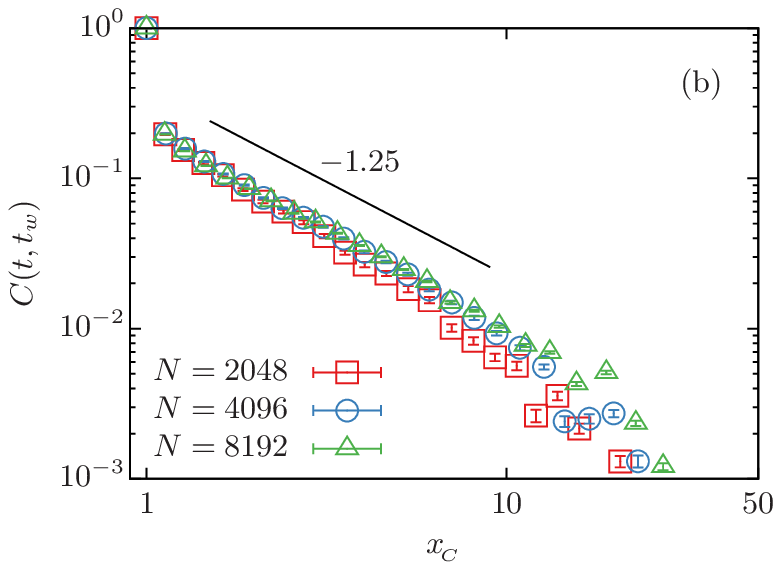}
 \caption{$C(t,t_w)$ for different $N=2048,4096$, and $8192$ for (a) Model\ I and (b) Model\ II quenched to $T_q=1.5$ with $t_w=10^3$ fixed against $x_{\!_C}=\ell(t)^3/\ell(t_w)^3$.}
\label{AgingDirect}
 \end{figure}
  \begin{figure}[t!]
  \includegraphics{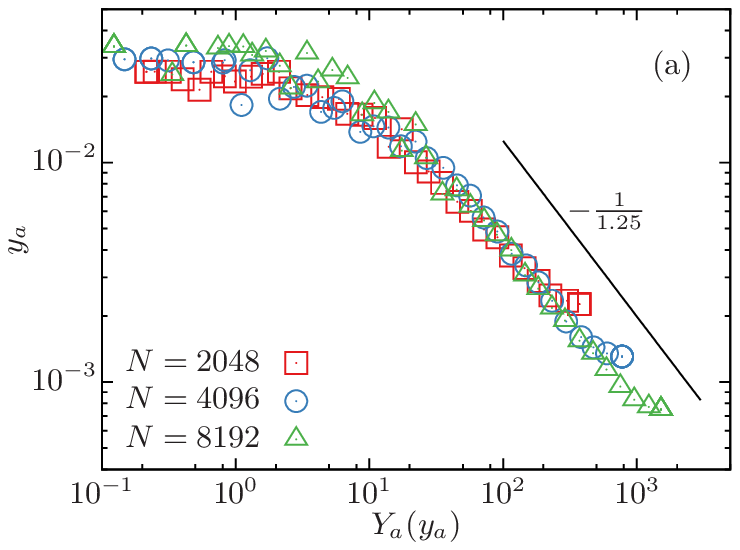}
  \includegraphics{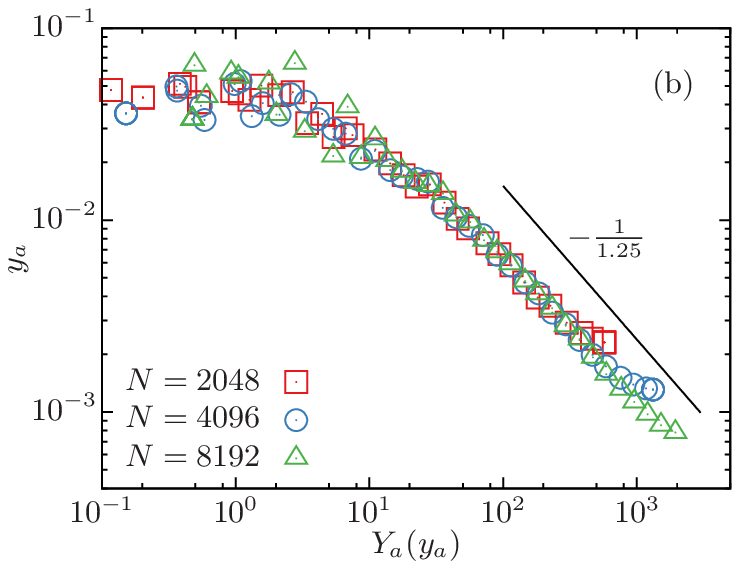}
 \caption{Finite-size scaling plot of $C(t,t_w)$ for $N=2048,4096$, and $8192$ for quench temperature $T_q=1.5$ with $t_w=10^3$ fixed for (a) Model\ I and (b) Model\ II. 
 The scaling behavior $y_a \sim Y_a^{-1/\lambda_C}$ is reasonable well observed for $\lambda_C=1.25$.}
\label{AgingFSS}
 \end{figure}
 \par
 Our next task is to have a measure of the dynamic exponent $\lambda_C$ governing the scaling \eqref{aging_scaling}. In Figs.\ \ref{Figtbytw}(e) and (f), 
 the scaling plots show that the data for $x_{\!_C}\gg 1$ is consistent with the continuous line having slope $-1.25$. In this regard, when the numerically precise 
 value \cite{Clisby2010} of $\nu$ is inserted in the bound \eqref{poly-bound} for $\lambda_C$, one gets 
 \begin{equation}
 0.762791 \leq \lambda_C \leq 1.525582.
 \label{LambdaRange}
 \end{equation}
 The obtained $\lambda_C=1.25$ from Figs.\ \ref{Figtbytw}(e) and (f) for both Model\ I and Model\ II thus not only follows the general theoretically bound \eqref{poly-bound}, but 
 seems to be in agreement with the numerical estimates obtained for the off-lattice model in Refs.\ \onlinecite{majumder2016evidence,majumder2016jpcm,majumder2017kinetics}. 
 \subsubsection{Finite-Size Scaling Analysis}
 To further substantiate the numerical estimate of $\lambda_C$, we call for a finite-size scaling analysis using data for three different system sizes shown in 
 Figs.\ \ref{AgingDirect}(a) and (b) respectively for Model\ I and Model\ II. Note that here we have used a fixed $t_w$ ($=10^3$ MCS), hence the onset of finite-size effects 
 (the downward tendency of the data) occurs earlier for smaller $N$. 
 We do a finite-size scaling analysis based on the scaling 
 form \eqref{aging_scaling} by writing down our finite-size scaling ansatz as 
   \begin{equation}
  NC(t,t_w)=Y_a(y_a),
 \end{equation}
where for the judicial choice of the scaling variable 
\begin{equation}
 y_a=x_{\!_C}(NA_C)^{-1/\lambda_C}
\end{equation}
one gets $Y_a \sim y_a^{-\lambda_C}$, i.e., $y_a \sim Y_a^{-1/\lambda_C}$. In our exercise, we fix $A_C=1$ and by varying $\lambda_C$ obtain reasonable data collapse 
for $\lambda_C=1.25(5)$, consistent with the obtained value for different waiting times $t_w$. In Figs.\ \ref{AgingFSS}(a) and (b) we show the representative plots 
for the finite-size scaling exercise for Model\ I and Model\ II, respectively, with $\lambda_C=1.25$. Both the models show reasonable quality of collapse of data 
and consistent behavior with $y_a \sim Y_a^{-1/\lambda_C}$. Note that here the crossover to the finite-size affected limit for smaller values of $Y_a$ from the scaling regime 
occurs rather gradually in contrast to the corresponding picture in an off-lattice model. \cite{majumder2016evidence}

 \subsubsection{Temperature-Dependent Scaling of Aging}
\begin{figure}[t!]
  \includegraphics{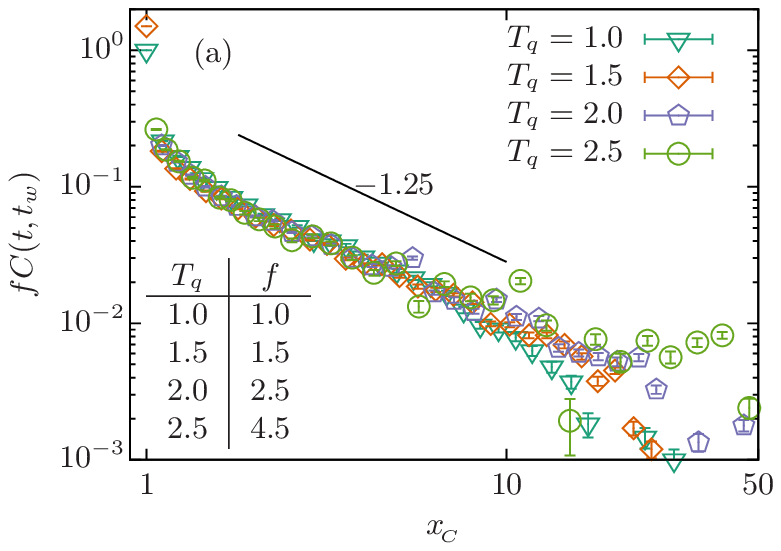}
  \includegraphics{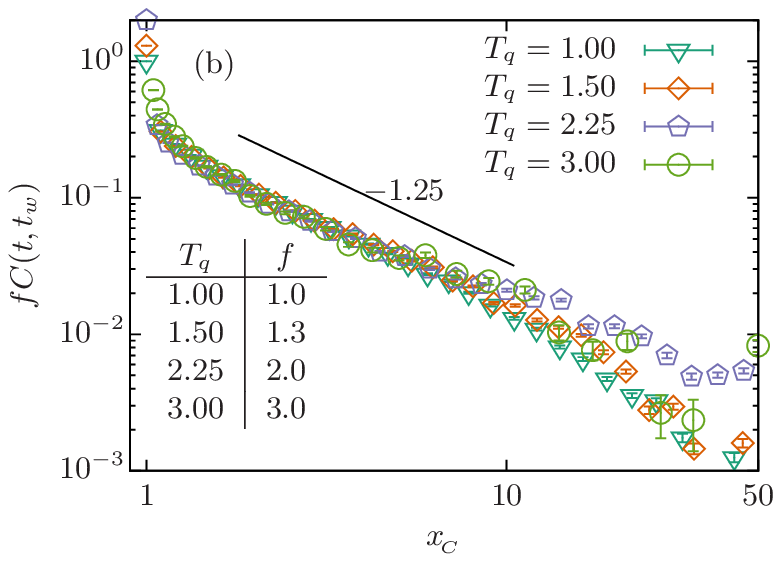}
 \caption{Plots to show the scaling of $C(t,t_w)$ with respect to $x_{\!_C}=\ell(t)^3/\ell(t_w)^3$ at different quench temperatures $T_q$ for (a) the fixed bond Model\ I and (b) the fluctuating bond Model\ II with $N=4096$, 
 for fixed $t_w \approx 10^3$. 
 To obtain the data collapse, the two-time correlation functions were multiplied by a growth amplitude dependent factor $f$, whose values are quoted in the figure.}
\label{AgingTemp}
 \end{figure}

 This universal nature of $\lambda_C$, irrespective of the details of models, thus urges us to check its robustness in the present models at different quench 
 temperatures, especially considering our observation of a rather non-universal nature of the other dynamic exponent $\alpha_c$ for Model\ I. In Fig.\ \ref{AgingTemp}, we 
 show for the chain of length $N=4096$ the temperature effect on the behavior of the autocorrelations $C(t,t_w)$ as a function of $x_{\!_C}$ for fixed waiting time $t_w=10^3$. There the 
 $y-$axis has to be multiplied by a factor $f=A_C(T_q=1)/A_C(T_q)$, similar to the metric factors in Fig.\ \ref{TempMod2}, to make them 
 collapse onto a single master curve. This master-curve behavior at different $T_q$ implies that, in contrast to the 
 cluster growth exponent, both models are found to be following the same power-law decay of the autocorrelations at different $T_q$. This further 
 strengthens the dynamic universal behavior of the aging exponent $\lambda_C$ concerning the collapse of a polymer.

\section{Conclusion}
\label{conclusion}
We have presented results from the kinetics of collapse for a lattice homopolymer in dimension $d=3$ using two different 
bond types, that is with fixed and flexible bonds. The cluster growth that occurs during the collapse is a scaling 
phenomenon as observed from the scaling of the two-point equal-time correlation function. However, for Model\ I with 
fixed bonds the growth of the clusters appears to be strongly dependent on the quench temperature. Similar observations were 
observed for ordering phenomena in disordered magnets.\cite{komori1999numerical,mouritsen1985temperature,paul2004domain,paul_PRE2005,Paul2007} 
On the other hand, Model\ II where 
the bonds have the flexibility of switching lengths between edges and diagonals of the lattice produces a much weaker dependence of the growth 
on the quench temperature. In fact, for moderately high temperatures, we show that the cluster growth can be 
described by an universal finite-size scaling function, pretty much like an off-lattice model. The growth exponent $\alpha_c \approx 0.62$
which we have estimated for such a description, however, is much smaller than that was observed for the off-lattice model 
\cite{majumder2015cluster,majumder2017kinetics} which exhibits a linear growth. This we feel is attributed to the topological constraint one encounters in a
lattice model, a fact, though expected, needs still to be verified. In this regard, it would be worth investigating 
the kinetics in the bond-fluctuation model.\cite{carmesin1988bond} In equilibrium it has 
been shown that it produces the correct dynamical picture and one must expect that the topological constraints can be overcome 
more easily, which may lead to a much more universal picture of the cluster growth.
\par
Regarding the other aspect of the kinetics of collapse, i.e., aging,  both the models produced a rather universal 
picture independent of quench temperature. Although the absence of scaling for the autocorrelation function with respect to $t/t_w$ 
suggested the presence of subaging, we have shown that the simple aging behavior is realized when one observes the scaling with respect to 
the ratio of growing cluster sizes, i.e. $C(t,t_w) \sim \left[\ell(t)^3/\ell(t_w)^3\right]^{-\lambda_C}$. We also show that the nonequilibrium autocorrelation exponent $\lambda_C$, 
governing such scaling is independent of the model as well as the quench temperature, and the observed value of $\lambda_C=1.25(5)$ not 
only follows the general bound $(\nu d-1) \le \lambda_C \le 2(\nu d-1)$ but also matches perfectly with the corresponding exponent from 
an off-lattice simulation. \cite{majumder2017kinetics,majumder2016evidence,majumder2016jpcm}
For ordering ferromagnets in dimension $d=2$ the nonequilibrium order-parameter autocorrelation exponent has been found to be $\lambda_C \simeq 5/4$,
\cite{fisher1988nonequilibrium,humayun1991non,liu1991nonequilibrium,henkel2001aging,henkel2004two,lorenz2007numerical,midya2014aging} 
matching with the value found in our present study for polymer collapse in lattice models and in Refs.\ \onlinecite{majumder2016evidence,majumder2017kinetics} for a continuum formulation, both in $d=3$ dimensions.
Due to the dimensional differences between the systems we feel that this could be a mere coincidence and further investigation is required to confirm any true non-trivial relationship.
\par
To conclude, we have shown how the methodologies popular in studies of ordering dynamics of spin models can well be used 
to understand the kinetics of polymer collapse. Hence, we strongly believe that by using a similar framework one 
could provide new insights in the mechanisms of other macromolecular conformational transitions such as collapse or 
folding of proteins or peptides using the HP model.\cite{dill1985theory,bachmann2004thermodynamics} Furthermore, it would also be interesting to check the 
validity of the various scaling laws discussed here for bulk polymers in lower dimension $d=2$ and for proteins in quasi-two-dimensional geometry, e.g., 
for macromolecules adsorbed on a substrate.

\begin{acknowledgments}
The project was funded by the Deutsche Forschungsgemeinschaft (DFG) under Grant Nos. JA 483/31-1 and JA 483/33-1, 
and further supported by the Leipzig Graduate School of Natural Sciences ``BuildMoNa,''
the Deutsch-Franz\"osische Hochschule (DFH-UFA) through the Doctoral College ``${\mathbb L}^4$'' under Grant No.\ CDFA-02-07, and the EU Marie Curie IRSES network DIONICOS under Contract No. PIRSES-GA-2013-612707.
\end{acknowledgments}

\end{document}